\documentclass[10pt]{article} 
\usepackage{amsfonts,amssymb,slashed,makeidx,latexsym,setspace}
\usepackage{graphicx,graphics,floatflt,amssymb,epsf,rotate,subfigure} 
\usepackage{times,cite,color}
\usepackage{multirow}
\usepackage{hyperref}
\begin{document}
\begin{flushright}
SINP/TNP/2012/01
\end{flushright}


\begin{center}
{\Large \bf Naturally split supersymmetry} \\
\vspace*{1cm} \renewcommand{\thefootnote}{\fnsymbol{footnote}} { {\sf
    Gautam Bhattacharyya${}^1$} and {\sf Tirtha Sankar Ray${}^{2}$}}
\\
\vspace{10pt} {\small ${}^{1)}$ {\em Saha Institute of Nuclear
    Physics, 1/AF Bidhan Nagar, Kolkata 700064, India} \\ ${}^{2)}$
  {\em Institut de Physique Th\'eorique, CEA-Saclay, F-91191
    Gif-sur-Yvette Cedex, France}}
\normalsize
\end{center}

\begin{abstract}
  Nonobservation of superparticles till date, new Higgs mass limits
  from the CMS and ATLAS experiments, WMAP constraints on relic
  density, various other low energy data, and the naturalness
  consideration, all considered simultaneously imply a paradigm shift
  of supersymmetric model building. In this paper we perform, for the
  first time, a detailed numerical study of brane-world induced
  supersymmetry breaking for both minimal and next-to-minimal
  scenarios. We observe that a naturally hierarchical spectrum emerges
  through an interplay of bulk, brane-localized and quasi-localized
  fields, which can gain more relevance in the subsequent phases of
  the LHC run.

\end{abstract}

\setcounter{footnote}{0}
\renewcommand{\thefootnote}{\arabic{footnote}}

\noindent {\bf Introduction:}~ With no sign of supersymmetry at the
CERN Large Hadron Collider (LHC) so far, even after the accumulation
of $\sim$ 5/fb data in the CMS and ATLAS experiments each, it is time
to reflect on those supersymmetric models which ($i$) can evade easy
detection at the early LHC run at 7 TeV \cite{Baer:2011ec}, ($ii$) can
solve problems related to large flavor changing neutral currents and
CP violation \cite{Gabbiani:1996hi}, ($iii$) can give sufficient relic
abundance of dark matter consistent with the WMAP data, and ($iv$) can
still manifest in a later phase of LHC at 14 TeV with more
luminosity. A minimal supersymmetric model (MSSM) spectrum like the
following can do the job: light Higgsinos (around a TeV), and heavy
other superpartners (few to several TeV squarks/sleptons, with a
relatively light stop, and super-heavy gauginos).  How natural is such
a spectrum?  Although a small Higgsino mixing parameter $\mu$ is
encouraging from the naturalness consideration, it still requires
fine-tuning to keep the quantum correction to the Higgs soft mass
under control. A generic expression for this correction is given by
$\Delta m^2 \sim (c/16\pi^2) m_{\tilde{q}}^2 \ln(M_S/M_Z)$, where $c$
is an order one coefficient for third generation and small for the
first two generation matter fields, $M_S$ is the messenger scale at
which supersymmetry is broken. The gluino contributes at the two loop
level, so the naturalness sensitivity to gluino mass is
small. Admittedly, the LHC data could not so far directly constrain
the third generation squarks/sleptons, but in most of the mediation
mechanisms the scalar masses of different generations are related.  As
LHC gradually pushes $m_{\tilde{q}}$ to higher values, naturalness
would prefer a relatively low $M_S$ (than the usual high scales
preferred by gravity or even by gauge mediation).  Here we take up a
class of 5d scenarios introduced some years back \cite{Diego:2006py}
where supersymmetry breaking proceeds via Scherk-Schwarz (SS)
mechanism \cite{Scherk:1979zr,pomarol} attributing improved
naturalness. However, nonobservation of the Higgs boson to date and
the WMAP relic density abundance cannot be simultaneously explained
within this context, and additionally, the superparticle spectra are
pushed beyond the reach of LHC.  We incorporate a few conceptual
inputs to resurrect a theoretically well-motivated framework that can
address all the current issues. Here gauge fields propagate in the
bulk and some (or all) matter fields are localized (with the Higgs
quasi-localized) at one of the branes. Supersymmetry is broken in the
bulk by SS mechanism through twisted boundary conditions, or
equivalently, by the vacuum expectation value (vev) of a radion living
in the bulk \cite{Marti:2001iw}.  We get a naturally split spectrum
where the bulk gauginos are ${\cal{O}}(10)$ TeV, while brane-localized
squarks/sleptons' masses are loop suppressed.  The soft masses are
gennerated at the scale $M_S$ itself, and $M_S \sim {\cal{O}}(10)$ TeV
implies a gain of a factor of $\sim$ 7 compared to mSUGRA in the
naturalness parameter \cite{hep-ph/0310137}. We scan over a wide range
of the model parameters to make our key observations as model
independent as possible. Adding an extra gauge singlet superfield,
quasi-localized near a brane, helps recover some parameter space lost
earlier to collider and cosmological data, and produce a lighter
spectrum with a possibility of enhanced visibility at a later phase of
the LHC run.

\noindent {\bf Supersymmetry breaking and soft scalar masses:}~ A 5d
$N=1$ vector supermultiplet can be decomposed from a 4d perspective
into a vector multiplet ${\mathcal{V}}(x,y)\supset
A_{\mu}(x,y),\lambda_1(x,y)$ and a chiral multiplet $\Phi(x,y) \supset
\phi(x,y), \lambda_2(x,y)$ in the adjoint representation of the gauge
group. Here, $A_\mu$ is the 4-vector gauge field, $\lambda_i (i=1,2)$
are gauginos, and $\phi\equiv (\Sigma + iA_5)/2$, where
$\Sigma$ is the 5d real scalar and $A_5$ is the 5th component of the
5-vector gauge field. The metric is given by $ds^2 = \eta_{\mu\nu}
dx^\mu dx^\nu + R^2 dy^2$, when the 5th coordinate is compactified on
$S^1/Z_2$ with a radius $R$.  The gauge invariant action of bulk
vector superfields coupled to a radion is given by \cite{Marti:2001iw}
\begin{equation}
 S^5_{\rm gauge} = \int d^4x ~dy \left[ \frac{1}{4 g_5^2} \int d^2
   \theta ~T~W^{\alpha} W_{\alpha} + {\rm h.c.} + \int d^4 \theta
   ~\frac{2}{g_5^2}~\frac{1}{T+ \bar T}~ {\left( {\partial}_y
     \mathcal{V} - \frac{1}{\sqrt{2}} \left( \Phi + \overline{\Phi}
     \right) \right)}^2 \right],
\label{lagra3}
\end{equation}
where $W^{\alpha}(x,y)$ is the field strength chiral superfield
corresponding to ${\mathcal{V}}(x,y)$. We can write $\langle T
\rangle=R+\theta^2 2 \omega$, where $\omega$ is the supersymmetry
breaking parameter. The mass spectrum of the component fields is given
by
\begin{equation}
 {\cal L}_{\rm gauge} = \frac{1}{R} \omega\lambda^{1\;(0)}\lambda^{1\;(0)} +
 \frac{n^2}{R^2}(A_{\mu}^{(n)}A^{\mu~(n)} + |\Sigma^{(n)}|^2) +
 \frac{1}{R}\left ( \lambda^{1\,(n)}\, \lambda^{2\,(n)} \right )
 \pmatrix{ \omega & n \cr n & \omega\cr }\left (\matrix{
   \lambda^{1\,(n)}\cr \lambda^{2\,(n)} \cr }\right ).
\label{massspc}
\end{equation}
Thus at the zero mode level we have a superfield $\mathcal{V} \supset
(A_{\mu}, \lambda_1)$ whose gauge component remains massless while its
gaugino acquires a Majorana mass $\omega/R$, where the supersymmetry
breaking parameter $\omega$ can be viewed as a twist in the ${\rm
  SU(2)}_R$ space of which $(\lambda_1, \lambda_2)$ is a doublet. Each
Kaluza-Klein (KK) mode consists of massive gauge bosons
$A_{\mu}^{(n)}$ and a real scalar $\Sigma^{(n)}$ each having masses of
the order of $n^2/R^2$ (the other real component is eaten up by the KK
gauge boson of the same level). Besides, there are two towers of
Majorana fermions $(\lambda^{1\,(n)}\pm \lambda^{2\,(n)})$ with masses
$|n\pm \omega|/R$.
The masses of the brane-localized ($y=0$) squarks/sleptons are
vanishing at tree level, and are generated at one-loop by gauge
interactions \cite{pomarol},
\begin{equation}   
m^2_{\widetilde \varphi}= \frac{g^2C_2({\widetilde \varphi})}{4\pi^4} \left[\Delta
  m^2(0)-\Delta m^2(\omega)\right]\, ,
\label{gaugecont}   
\end{equation} 
where $\Delta m^2(z) \equiv \frac{1}{2R^2}\left[Li_3(e^{i2\pi
    z})+Li_3(e^{-i2\pi z})\right] $, with
$Li_n(x)\equiv\sum^{\infty}_{k=1}{x^k}/{k^n}$.  Here, $C_2({\widetilde
  \varphi})$ is the quadratic Casimir of the
$\widetilde\varphi$-representation under the SM gauge group. It is
important to note that if the Higgs fields are localized, they receive
only positive contributions from the gauge multiplets.

\noindent {\bf Electroweak Symmetry Breaking (EWSB):}~
The Higgs soft masses also receive brane-localized top-stop
(bottom-sbottom) loop contributions, given by \cite{Diego:2006py}
\begin{eqnarray}
m_{H_u}^2=\frac{3 y_t^2}{8 \pi^2}m_{\widetilde{t}}^2 \, \log
\frac{m_{\widetilde{t}}^2 R^2}{\omega}, ~~ m_{H_d}^2=\frac{3 y_b^2}{8
  \pi^2}m_{\widetilde{b}}^2 \, \log \frac{m_{\widetilde{b}}^2
  R^2}{\omega}. 
\label{higg2}
 \end{eqnarray}
 This contributions in Eq.~(\ref{higg2}) can by itself trigger EWSB,
 but being a two-loop effect (since $m_{\tilde{t},\tilde{b}}$ are
 generated at one-loop) finds it hard to overcome a much larger
 one-loop positive contribution to $m_{H_u}^2$ as given by
 Eq.~(\ref{gaugecont}).  A resolution to this is to keep the $H_u$ and
 $H_d$ hypermultiplets quasi-localized near the $y=0$ brane
 \cite{Diego:2006py}. The advantage of quasi-localization is two-fold:
 ($i$) a bulk tachyonic mass can be generated using boundary
 conditions, and ($ii$) its mass is controlled by the supersymmetric
 mass $M$ (and not $1/R$) by which quasi-localization occurs,
 involving a suppression factor $\epsilon = \exp(-\pi MR)$. As a
 result, the bulk tachyonic mass and the one-loop mass of
 Eq.~(\ref{gaugecont}) can be of the same order, and a cancellation
 between them allows the two-loop contribution of Eq.~(\ref{higg2})
 dominate and trigger EWSB. The up- and down-type Higgs
 hypermultiplets form a doublet of a ${\rm SU(2)_H}$ global symmetry
 of the Lagrangian.  To generate a tachyonic mass one imposes suitable
 boundary conditions which create a twist $(\tilde{\omega})$ in that
 basis.  The action of the bulk Higgs hypermultiplets coupled to the
 bulk vector and radion superfields can be written as
 \cite{Diego:2006py},
\begin{eqnarray}
S^5_{\rm Higgs} &=& \int d^4x ~dy \left[ \int
  d^4\theta\ \frac{T+\bar T}{2} \left\{\bar{\mathcal H} \,
  e^{(\tau_a {\mathcal{V}}^a)} \, \mathcal H + \mathcal H^{c}\,
  e^{(-\tau_a{\mathcal{V}}^a)}\, \bar{\mathcal H}^c\right\}
  \right.\nonumber \\ &-&\left. \int d^2\theta \left\{\mathcal H^c
  (\partial_y-\mathcal M T -\frac{1}{\sqrt{2}}\Phi) \mathcal H+
  \delta(y-f)\frac{1}{2} \mathcal H^c[1 + \vec s_f\cdot\vec
    \sigma]\mathcal{H} + {\rm h.c.}\right\} \right] \, , 
\label{bulklag}
\end{eqnarray}
with hypermultiplet indices suppressed. The mass matrix $\mathcal M$
is hermitian and non-diagonal in ${\rm SU(2)_H}$ basis, given by
\begin{eqnarray}
 \mathcal M &=& M'+ M~ p^{\alpha} \sigma_{\alpha} = a_0/R +
 (a/R)~ p^{\alpha} \sigma_{\alpha} \, , 
\label{bulkmass}	  
\end{eqnarray}
where $\alpha$ in the ${\rm SU(2)_H}$ index, and $a_0$ and $a$ are
dimensionless order one coefficients.  Here $\vec s$ and $\vec p$ are
unit vectors in the ${\rm SU(2)_H}$ space, and $(1\pm\vec s_f\cdot\vec
\sigma)$ projects out a linear combination of the two ${\rm SU(2)_H}$
doublet whose wave function goes to zero at the boundary.  A
misalignment between $\vec s_0$ and $\vec s_\pi$ causes different
field combinations to survive at the two boundaries and creates a
supersymmetry preserving twist angle $\tilde{\omega}$, given by
$\cos(2\pi \tilde \omega)=\vec s_0\cdot \vec s_{\pi}$.

The bulk mass term $M'$ in Eq.~(\ref{bulkmass}) was set to zero in
\cite{Diego:2006py} to avoid the occurence of linearly divergent
$(\sim M'\Lambda)$ Fayet-Iliopoulos (FI) term. Since 5d theories are
inherently non-renormalizable and the cutoff in our kind of scenario
is rather low, we consider putting $a_0=0$ is unnecessarily
over-restrictive. We relax this constraint and turn on a small value
of $a_0$ to allow the most general form of the bulk mass.  We shall
highlight its advantages in this paper. The soft masses of the
quasi-localized up/down-type Higgses can be written as
\begin{eqnarray}
  m_{H_{u/d}}^2 \sim M^2 \sin^2(\pi\omega) (1-\tan^2(\pi\tilde
  \omega))\ \epsilon_{\mp}^2 \, , 
\label{higg3}
\end{eqnarray}
where $\epsilon_{\mp} = e^{-\pi ( a \mp a_0)} \ll 1$.  For
$\tilde\omega>1/4$ it is possible to get a tachyonic soft mass-square,
while for $\epsilon \sim 10^{-2}$ the tachyonic terms can effectively
cancel the positive contribution from the gaugino loops of
Eq.~(\ref{gaugecont}). Note that to arrange such a cancellation we
simply have to put $a \sim {\cal{O}} (1)$, thus we do not pay any
serious fine-tuning price.

\begin{figure*}
\begin{minipage}[t]{0.47\textwidth}
\vspace{.1 cm}
\begin{center}
\includegraphics[width=.9\textwidth, keepaspectratio]
  {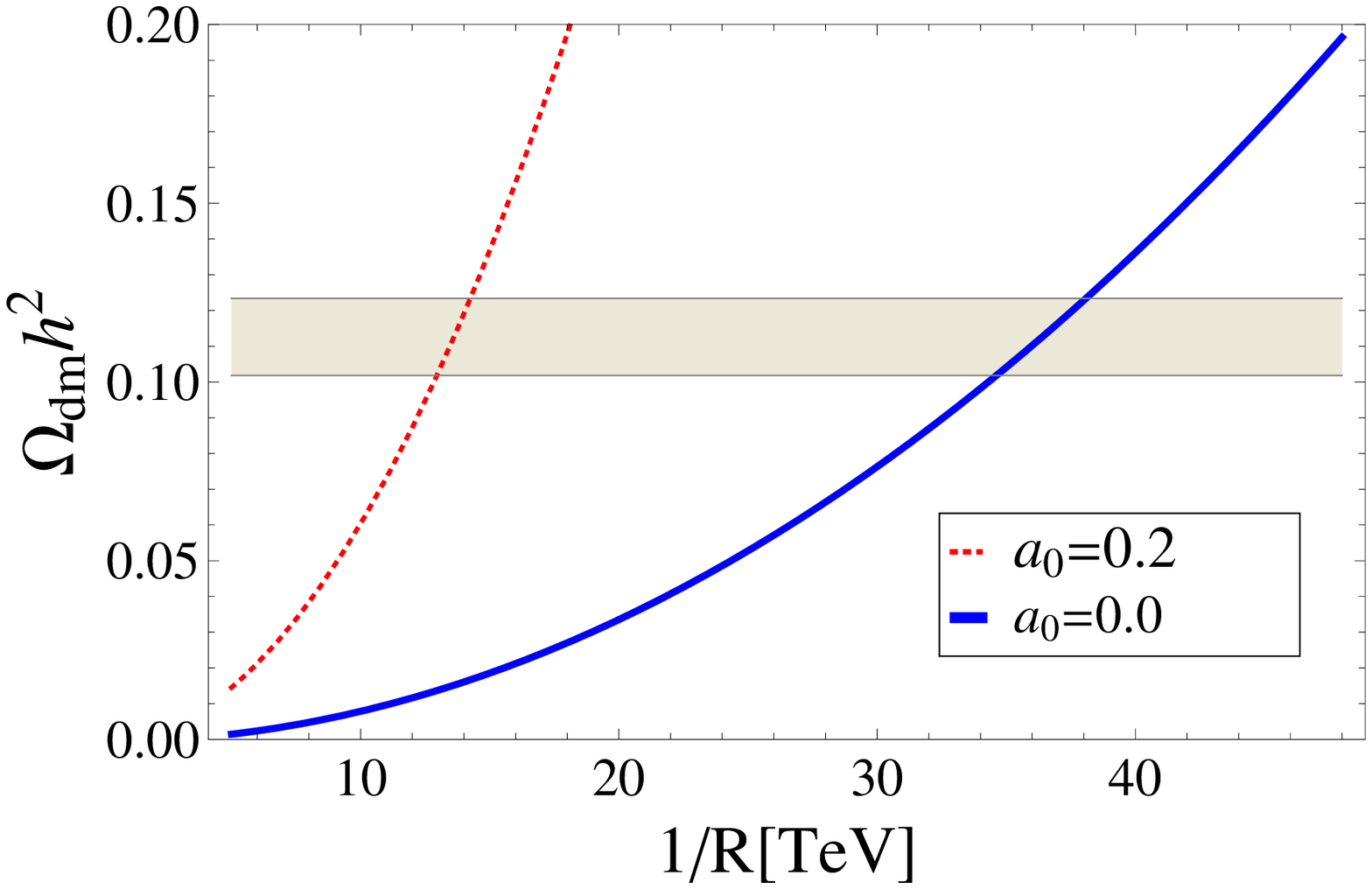}
  \caption[]{\em \small The dark matter density for $a=1.65,
    \omega=0.45$ and $\tilde{\omega}=0.35$. The shaded region
    corresponds to the $3\sigma$ allowed region from WMAP
    \cite{Komatsu:2010fb}.}
\label{fig1}
\end{center}
\end{minipage}
\hspace{7mm}
\begin{minipage}[t]{0.47\textwidth}
\begin{center}
\includegraphics[width=.5\textwidth,angle=270, keepaspectratio]{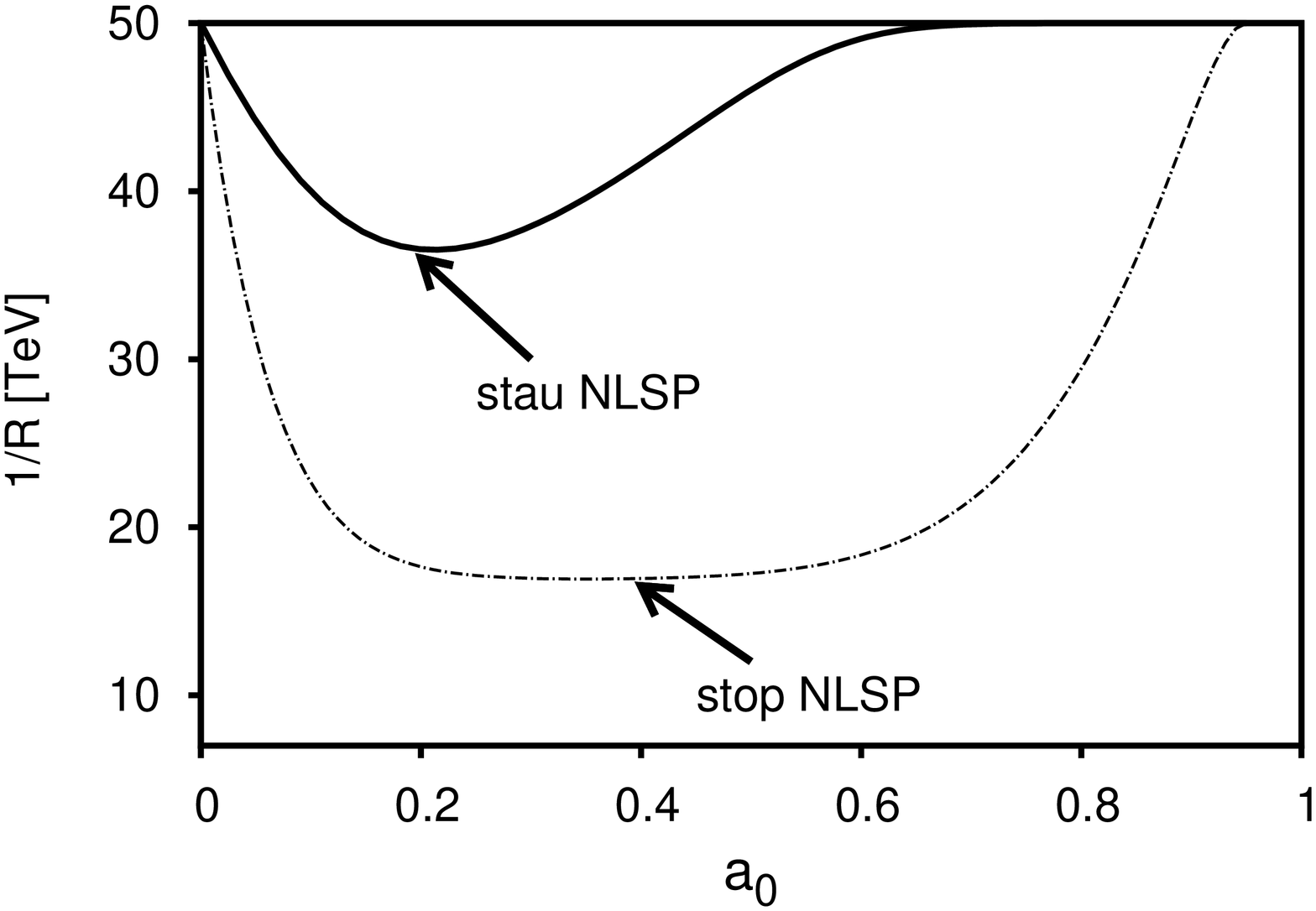}
\vspace{.7cm}
\caption[]{\em \small The lower limit of $R^{-1}$ from all data for
  two different scenarios.}
\label{fig2}
\end{center}
\end{minipage}
\end{figure*}
\noindent {\bf The parameter space of the model:}~ In Fig.~\ref{fig1}
we demonstrate that with $a_0 \neq 0$ the relic density attains the
WMAP allowed value for a relatively smaller value of $R^{-1}$.  A
nonzero $a_0$ increases the value of $\mu$ obtained from potential
minimization. The lightest supersymmetric particle (LSP) in our model
is {\em always} a Higgsino, and when the dark matter is Higgsino dominated
it turns out that $\Omega_{\rm DM} h^2 \simeq 0.09(\mu/{\rm TeV})^2$
for $\mu \gg M_Z$ \cite{Giudice:2004tc}. Consistency with the WMAP
data \cite{Komatsu:2010fb} thus allows a lighter spectrum for $a_0
\neq 0$.

\begin{figure*}[htbp]
\centering \subfigure
           {\includegraphics[width=.35\textwidth,keepaspectratio]
             {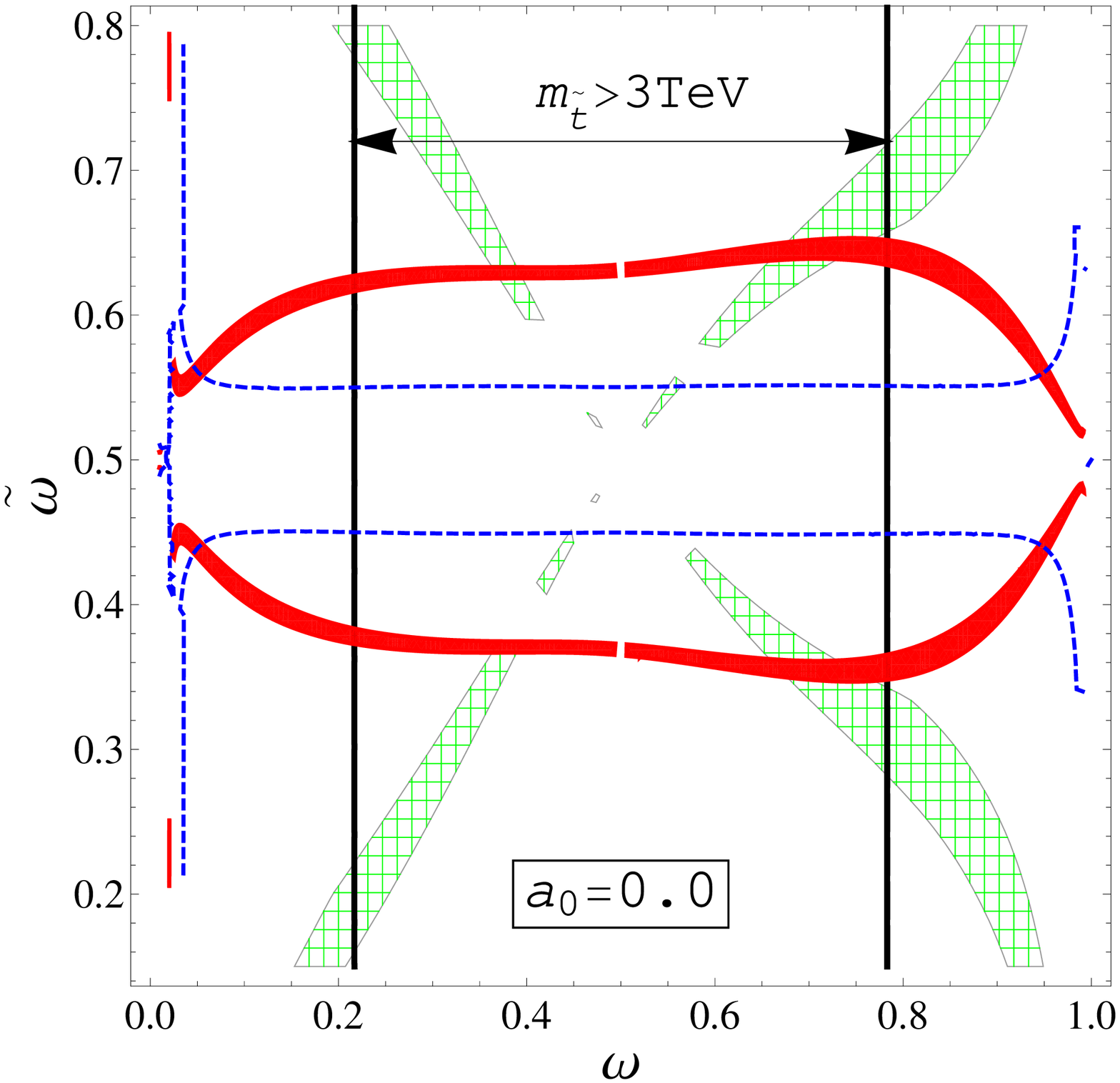}}
           \subfigure{\includegraphics[width=.35\textwidth,keepaspectratio]
             {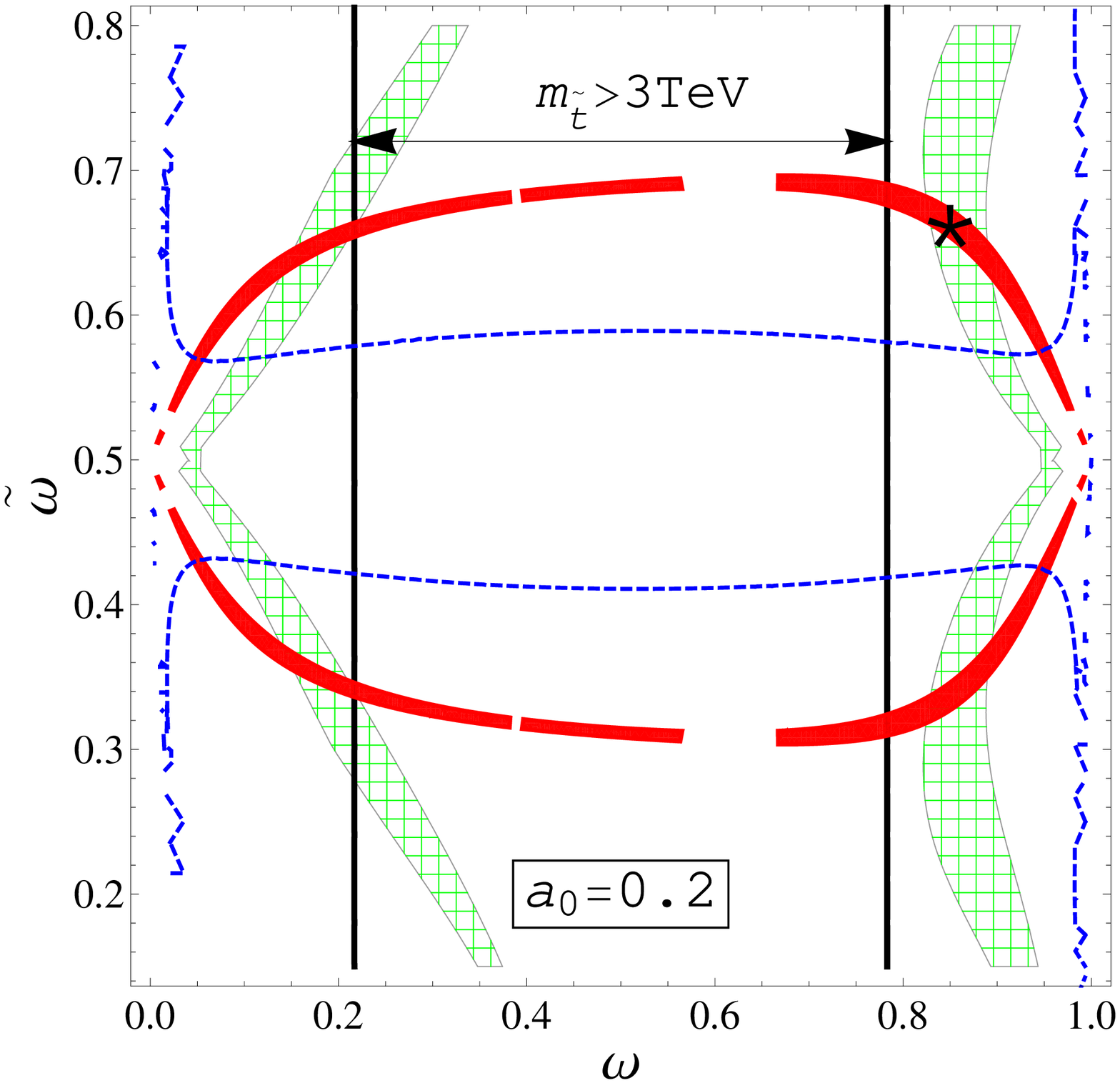}}
           \caption[]{\em \small Allowed/disallowed zone in the twist
             parameters space for $1/R=40$ TeV and $a=1.65$.  The
             green checkered region is compatible with EWSB and $115 <
             m_h < 127$ GeV.  The red shaded region is allowed by WMAP
             relic density. In between the dotted lines the stop
             becomes lighter than the lightest neutralino. For
             $a_0=0.2$ the region marked $(\ast)$ on the upper right
             corner maps to the parameter space where large charged
             tracks may be expected (see text).}
\label{fig3}
\end{figure*}
In Fig.~\ref{fig2} we display the lower limit of $R^{-1}$ as a
function of $a_0$ considering {\em all} data, especially the WMAP
relic density abundance ($0.1018<\Omega_{\rm DM} h^2<0.1234$)
\cite{Komatsu:2010fb}, the Higgs mass limits from CMS and ATLAS
experiments ($115 < m_h < 127$ GeV) \cite{Higgscombine}, and lower
limits on squarks/slepton masses set by Tevatron and LHC
\cite{LHCsusy}. For numerical estimates we have used the code {\sf
  micrOMEGAS} \cite{Belanger:2001fz}.  When all the three generation
matter fields are brane-localized, the lower limit on $R^{-1}$ is
around 35 TeV, which was 50 TeV for $a_0=0$.  The main source of this
constraint is the tension between the compatibility of EWSB occurrence
and the allowed range of $m_h$, which tends to make the stau lighter
than the Higgsino. However, if we keep $Q_3$ and $t_R$ localized at
$y=0$ brane but allow all other matter fields travel in the bulk, then
a stop (not a stau) becomes the next-to-lightest supersymmetric
particle (NLSP)\footnote{Nonuniversal localization of fermions in the
  bulk, motivated for explaining the fermion mass hierarchy
  \cite{ArkaniHamed:1999dc}, generally leads to dangerous FCNC and CP
  violating operators induced by tree level flavor violating couplings
  of KK-gluons with the SM fermions. For different localizations of
  the first two families the bound on the compactification scale
  arising from $\Delta M_K$ and $\epsilon_K$ is quite strong ($1/R ~>~
  5000$ TeV) \cite{Delgado:1999sv}. Since in our case the first two
  families reside in bulk having identical 5d bulk masses, and only
  the third family quarks are brane-bound, the corresponding operators
  are CKM suppressed and the bound is much weaker ($1/R ~>~ 4$ TeV)
  \cite{Delgado:1999sv}. Also, the first two generation squarks for
  this case are too heavy to create any flavor problem at one-loop
  level.}. In this stop NLSP case, as we see from Fig.~\ref{fig2}, the
WMAP constraint gets relaxed and the lower limit on $R^{-1}$ comes
down to 16 TeV.

In Fig.~\ref{fig3} we show the constraints in the plane of the twist
parameters $\omega$ and $\widetilde \omega$. The red shaded patches
are regions where our predicted relic density is consistent with WMAP
data. A nonvanishing $a_0$ shifts the overlap of these patches with
the green chequered zone (simultaneously satisfied by EWSB and the new
Higgs mass limits) to a region where the lighter stop weighs around 2
TeV.

\begin{figure*}
\begin{minipage}[t]{0.44\textwidth}
\begin{center}
\includegraphics[width=0.7\textwidth,angle=270,keepaspectratio]
{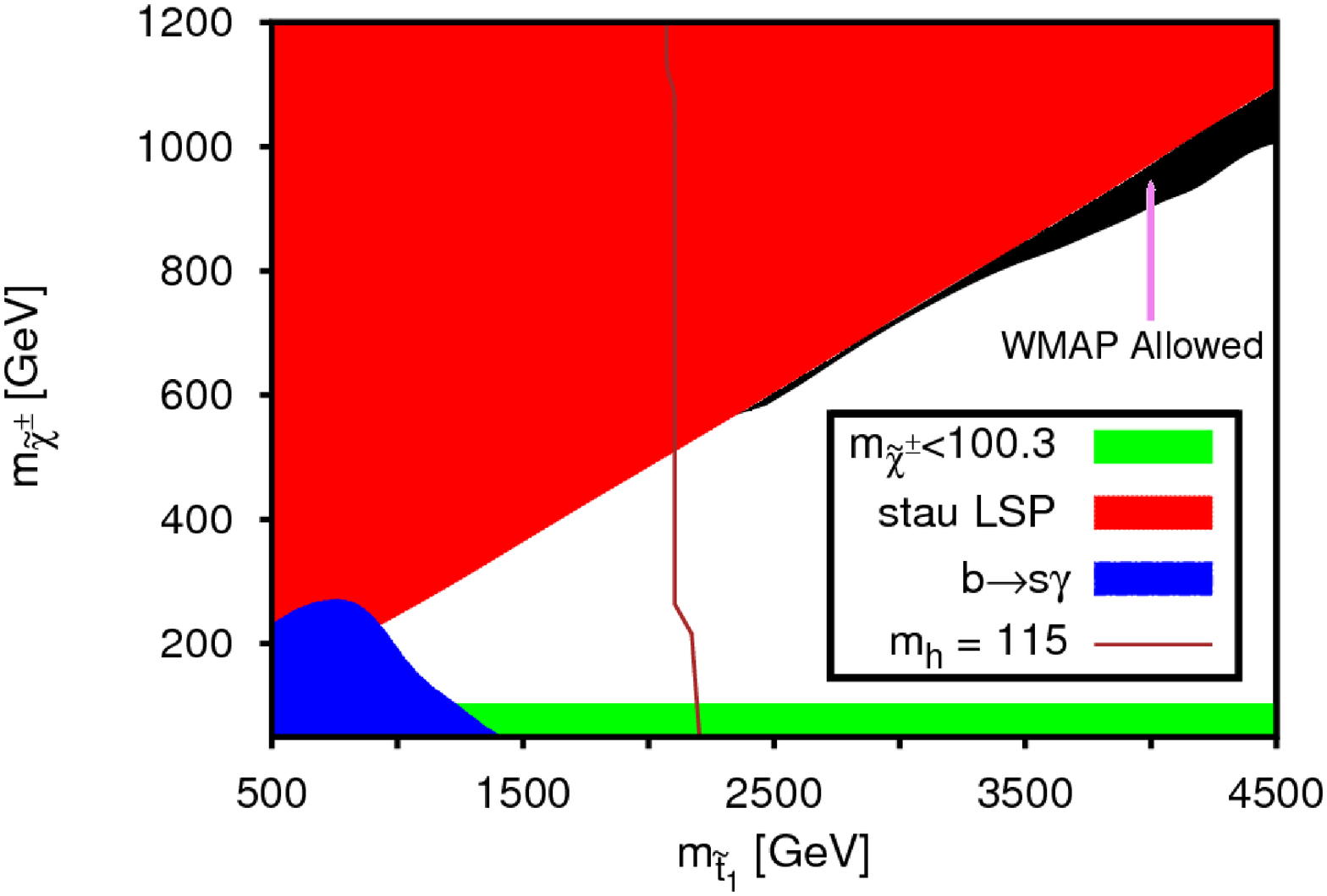}
\caption[]{\em \small Allowed/disallowed zone in the lightest stop-
  and chargino-mass plane. Only the black region is compatible with
  all data including WMAP.}
\label{fig4}
\end{center}
\end{minipage}
\hspace{7mm}
\begin{minipage}[t]{0.44\textwidth}
\begin{center}
\includegraphics[width=0.7\textwidth,angle=270,keepaspectratio]
{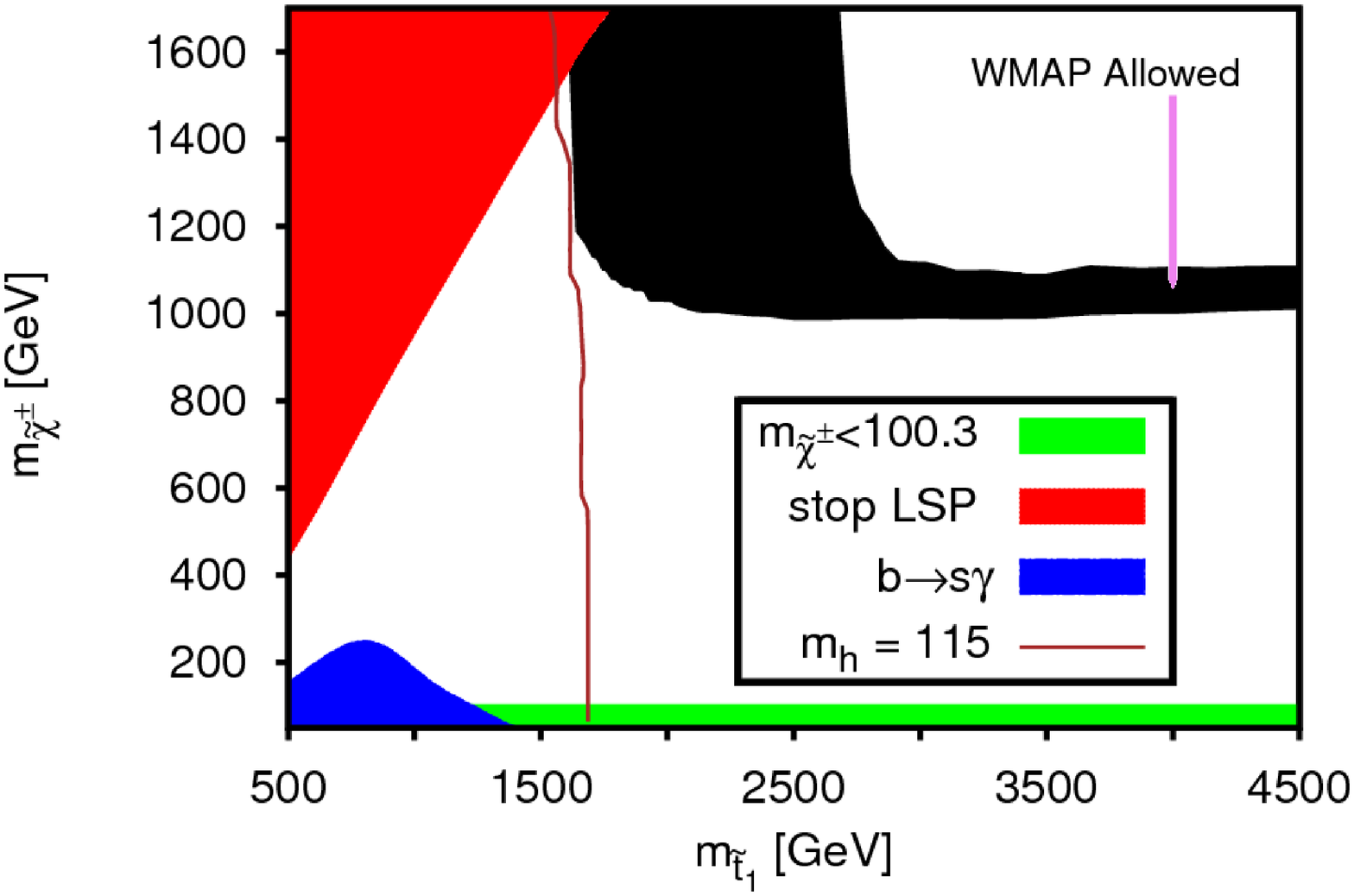}
\caption[]{\em \small Same as Fig.~\ref{fig4}, but only $Q_3$ and
  $t_R$ are brane-localized, i.e. when the stop is the NLSP.}
\label{fig5}
\end{center}
\end{minipage}
\end{figure*}

In Figs.~\ref{fig4} and \ref{fig5}, we plot the constraints in the
parameter space of the lighter stop mass (lightest colored sparticle)
and the lighter chargino mass, when all the model parameters of the
theory have been summed over in appropriate ranges. In Fig.~\ref{fig4}
all matter superfields are brane-localized, whereas in Fig.~\ref{fig5}
only $Q_3$ and $t_R$ are brane-localized.  In both cases $\tan\beta$
obtained from potential minimization varies between 3 and 15, and the
trilinear coupling $A_t$ is loop suppressed.  Being almost
Higgsino-like, the lighter chargino and the lightest neutralino are
highly degenerate $\sim \mu$, the degeneracy being mildly lifted by
radiative corrections.  A substantial part of the parameter space in
Fig.~\ref{fig4} is disfavored by a stau becoming an LSP.  In
Fig.~\ref{fig5}, however, where the stop is lighter than the stau, a
substantial part of the lost region is recovered.  We see that a stop
mass as light as 1.6 TeV is allowed in Fig.~\ref{fig5}, the main
constraint on it coming from the Higgs mass lower limit. There is a
substantial increase in the allowed territory (the black shaded
region) which satisfies all data mentioned earlier and also the
measurement of $(g-2)_{\mu}$ \cite{pdg}. The blue shaded region in
both Figs.~\ref{fig4} and \ref{fig5} is excluded by $b \to s \gamma$
at $3\sigma$ \cite{Barberio:2008fa}.  To make all these plots as model
independent as possible we have integrated over the model parameters
over the following range: $1/R \supset [0.5:50]~ {\rm TeV}, \omega
\supset [0:1], \tilde{\omega} \supset [0:1], a \supset [1:2]
~\mbox{and}~ a_0 \supset [0:1]$.  The lighter spectrum of
Fig.~\ref{fig5} mimics that of the `partially supersymmetric model'
explored in \cite{Gherghetta:2003wm}. In most of the allowed region of
Fig.~\ref{fig5} the fine-tuning is about 10\%. 

The near equality between $m_{\tilde\chi^\pm}$ and $m_{\tilde\chi^0}$
constitutes a characteristic signature of this scenario. Within the
allowed region of the model parameters, for $1/R = 40~(16)$ TeV, we
estimate $\Delta m_\chi \equiv m_{\tilde\chi^\pm} - m_{\tilde\chi^0}$
to lie in the range of 100 to 150 (300 to 400) MeV, which correspond
to decay length 1m to 10 cm ($\sim$ 0.5 cm) \cite{Gunion:1999jr}. It
is therefore not unexpected to observe a large charged track with
heavy ionization, which corresponds to the region marked $(\ast)$ in
Fig.~\ref{fig3}.

\noindent {\bf NMSSM using a quasi-localized singlet:}~ The
next-to-minimal supersymmetric models (NMSSM) offers quite a few
advantages \cite{Ellwanger:2009dp}: it solves the $\mu$ problem, it
can hide a Higgs boson under the cover of its singlet admixture, it
has a better WMAP compatibility through a mixed singlino-Higgsino dark
matter, etc.  We construct a brane-world NMSSM model by
quasi-localizing a gauge singlet with a supersymmetric mass $M$, like
what we did earlier for $H_u, H_d$ hypermultiplets.  We show that the
tachyonic mass of the singlet scalar indeed helps to generate its vev.

Dropping the Yukawa terms we write the superpotential and the soft
breaking part of the Lagrangian as,
\begin{equation}
 W \supset \lambda \mathcal{S} \mathcal{H}_u\cdot \mathcal{H}_d +
 \frac{1}{3}\kappa\,\mathcal{S}^3, ~~ -\mathcal{L}_{\rm soft} \supset
 m_S^2|S|^2 + \left( \lambda A_{\lambda} H_u\cdot H_d S +
 \frac{1}{3}\kappa A_{\kappa}S^3 + {\rm h.c.} \right).
\label{spnmssm}
\end{equation}
The vev $s$ of the singlet scalar $S$ is given by $\langle s\rangle
\simeq \frac{1}{4 \kappa} \left( -A_{\kappa} + \sqrt{A_{\kappa}^2 - 8
  m_{S}^2} \right)$, when $s \gg v_u, v_d$.  A nonvanishing $s$
therefore means either $A_{\kappa}>m_S^2$ or $m_S^2<0$.  Since in our
scenario $A_\kappa$ is very suppressed (see later), we stimulate the
$m_S^2<0$ option from brane-world dynamics. However, such an extreme
choice of parameters, namely, $s \gg (v_u, v_d)$, is meant only for
simple illustration.  While doing the full potential minimization we
do not make any such assumption and numerically check at every stage
that EWSB does occur and low energy observables are satisfied. The
bottom line is that by employing a tachyonic soft mass-square we can
arrange for a nonvanishing vev $s$ which leads to consistent
phenomenology.

To follow the same method for quasi-localization we employed for $H_u$
and $H_d$ we must introduce an ${\rm SU(2)_H}$ index to describe the
bulk gauge singlet hypermultiplet $\mathbb S$.  We write the multiplet
as $\mathbb S^\alpha=(S_i,\Psi_s,F_{s~i})^\alpha$, by splitting the
complex hypermultiplet into two real parts, using the label $\alpha$
for the ${\rm SU(2)_H}$ index and $i$ is the ${\rm SU(2)_R}$
index. One can introduce $\omega$ and $\tilde\omega$ exactly like
before.  It was shown in \cite{hep-ph/0505244}, though not
specifically in the NMSSM context, that for suitable boundary
conditions and for $\omega=\tilde\omega =1/2$, a tachyonic mass
$m_S^2= - 4 M^2 \exp( -\pi MR)$ can be generated for a singlet scalar
whose wavefunction peaks at $y=0$.  The values of $A_\lambda$ and
$A_\kappa$ are assumed to be zero at the scale $1/R$ and their values
at the weak scale can be computed from
\begin{eqnarray}
{d A_\lambda \over d t} &=& {1\over 16\pi^2}\left[ 6 A_t \lambda_t^2 +
  8 \lambda^2 A_\lambda + 4 \kappa^2 A_\kappa + 6 g_2^2 M_2 +
  \left({6\over 5}\right) g_1^2 M_1 \right] \Longrightarrow A_\lambda(M_W)
\sim .08\frac{\omega}{R} \, ; 
\nonumber\\ {d A_\kappa \over d t} &=& {12\over 16\pi^2} \left(
\lambda^2 A_\lambda + \kappa^2 A_\kappa \right) \Longrightarrow
A_\kappa(M_W) \sim .014\frac{\lambda^2\omega}{R} \, .
\label{nmssma}
\end{eqnarray}
From the full scalar potential minimization we fix $v_u$, $v_d$ and
$s$ and we are left with seven free parameters: $R^{-1}$, $a$, $a_0$,
$\omega$, $\tilde\omega$, $\lambda$, $\kappa$. For this NMSSM case we
can afford to set $a_0 = 0$. To obtain the spectrum and the various
constraints we use the package \texttt{NMSSMTools}
\cite{Ellwanger:2004xm} modified for split spectrum like ours and
linked to {\sf micrOMEGAS} \cite{Belanger:2001fz}. The key features
for a benchmark point $R^{-1} = 11$ TeV, $a=1.6$, $\omega=0.57$,
$\tilde\omega=0.66$, $\lambda=0.4$, $\kappa=0.06$ are the following:
($i$) $m_{h_1} \approx 59$ GeV and $m_{h_2} \approx 111$ GeV (this can
evade the LEP-2 bound), where the lighter of the two CP-even Higgs
states has a 99\% branching fraction of decaying into two CP-odd
states with a mass $m_{h_a} \simeq 9.4$ GeV; ($ii$) the dark matter is
the lightest neutralino with mass $\approx 56$ GeV with a large
singlino component ($\approx 0.93 \tilde S$); ($iii$)
$\Omega_{\tilde{\chi_1^0}} h^2 \approx 0.1$; ($iv$) ${\rm Br}(b \to s
\gamma) = 3.54 \cdot 10^{-4}$; ($v$) ${\rm Br}(B_s \to \mu^+ \mu^-) =
2.77 \cdot 10^{-9}$. Unlike in the MSSM scenario, $\Delta m_\chi$ is
much higher here (around 116 GeV for this particular case). Its
signals would be similar to as expected in the `light Higgsino-world
scenario' \cite{Baer:2011ec} but with enhanced cross section due to
larger splitting.

\noindent {\bf Conclusions:}~ By the end of 2011 supersymmetric model
building has entered a new era, where the conventional gravity or
gauge mediation models are feeling increasingly uncomfortable.  The
expected nature of superparticle spectrum is hierarchical. In this
paper we have done the first detailed numerical study of a general
class of brane-world inspired MSSM scenario and its NMSSM extension by
confronting all laboratory and cosmological data. Some characteristic
signatures are also mentioned. One of the highlights of this work is
an elegant implementation of NMSSM for the first time using SS
mechanism by exploiting the generation of tachyonic soft mass-square
using ${\rm SU(2)_H}$ rotation. In spite of its hierarchy such models
suffer less from naturalness problem because of the low messenger
scale at which supersymmetry is broken. This class of models is likely
to gain more relevance during 2012 and beyond.

\noindent {\bf Acknowledgments:}~ We thank G.v. Gersdorff, D. Das and
C.A. Savoy for valuable suggestions. GB acknowledges hospitality at
CEA, Saclay, during a visit where this work was initiated, and both
authors thank CERN Theory Division for hospitality during the final
stage of this work.  The work of TSR is supported by EU ITN, contract
"UNILHC" PITN-GA-2009-237920, the CEA-Eurotalents program and the
Agence Nationale de la Recherche under contract ANR 2010 BLANC 0413
01.



\end{document}